\documentclass[aps,prb,twocolumn,showpacs,superscriptaddress,10pt]{revtex4-1}

\usepackage{graphicx}
\usepackage{amssymb}
\usepackage[ansinew]{inputenc}

\begin{document}

\title{Screening of electron-phonon coupling in graphene on Ir(111)}

\author{M.~Endlich}\email{michael.endlich@tu-ilmenau.de}\affiliation{Institut für Physik, Technische Universität Ilmenau, D-98693 Ilmenau, Germany}

\author{A.~Molina-Sánchez}
\author{L.~Wirtz}
\affiliation{Physics and Materials Science Research Unit, University of Luxembourg, L-1511 Luxembourg}
\affiliation{Institute for Electronics, Microelectronics, and Nanotechnology (IEMN), CNRS UMR 8520, Dept.\ ISEN, F-59652 Villeneuve d'Ascq Cedex, France}

\author{J.~Kröger}\affiliation{Institut für Physik, Technische Universität Ilmenau, D-98693 Ilmenau, Germany}

\begin{abstract}
The phonon dispersion of graphene on Ir(111) has been determined by means of angle-resolved inelastic electron scattering and density functional calculations.  Kohn anomalies of the highest optical phonon branches are observed at the $\bar{\Gamma}$ and $\bar{\text{K}}$ point of the surface Brillouin zone.  At $\bar{\text{K}}$ the Kohn anomaly is weaker than observed from pristine graphene and graphite.  This observation is rationalized in terms of a decrease of the electron-phonon coupling due to screening of graphene electron correlations by the metal substrate.
\end{abstract}


\maketitle

\section{Introduction}

Since its discovery graphene has attracted particular attention owing to the extremely high charge carrier mobility, the ballistic electron transport at room temperature and the ambipolar electric field effect. \cite{kno_04,kno05a,kno05b,yzh_05}  To preserve the unique properties of graphene on surfaces as much as possible the preparation and characterization of single layers of graphene on metal surfaces has become an increasingly important research field. \cite{jwi_09}  Among the large variety of metal surfaces Ir(111) and Pt(111) are outstanding.  The graphene--surface distance is $\approx 340\,\text{pm}$ on Ir(111) \cite{cbu_11} and $\approx 330\,\text{pm}$ on Pt(111). \cite{psu_09}  These values represent the largest distances that have been reported for graphene on metal surfaces so far and imply a weak graphene--metal interaction.  Indeed, the characteristic electronic structure of graphene on Ir(111) is weakly affected since the Dirac cones at the $\bar{\text{K}}$ point of the surface Brillouin zone are shifted only slightly above the Fermi level. \cite{ipl_09}

The dynamical properties of carbon materials have been shown to severely influence their electron transport properties. For instance, electron scattering from optical phonons leads to a collapse of the ballistic electron transport in carbon nanotubes. \cite{zya_00,aja_04}  Likewise, the transport properties of graphene in the high-current limit are affected by the interaction between electrons and optical phonons. \cite{bar_09}  In addition, exploring phonon spectra of graphene may provide valuable information on the bonding of graphene with the substrate, \cite{from_13} the persistence of the Dirac cone, \cite{aal_10} and the electron-phonon coupling. \cite{spi_04}  Therefore, the precise knowledge of the phonon band structure of graphene on a metal surface is highly desirable for the understanding of the graphene--metal interaction and its underlying physics.  However, the dynamical properties of graphene on metal surfaces with weak interaction have scarcely been addressed so far.  Raman spectroscopy has been used to determine characteristic graphene phonon modes at $\bar{\Gamma}$ on two rotational variants of epitaxial, single-layer graphene on Ir(111). \cite{est_11}  The phonon dispersion relations of graphene on Pt(111) have been determined by electron energy loss spectroscopy along the $\bar{\Gamma}\bar{\text{K}}$ direction of the surface Brillouin zone. \cite{apo_12} The similarity with the phonon dispersion of graphite has been interpreted in terms of a weak graphene--Pt interaction.

One of the key signatures of electron-phonon coupling is the Kohn anomaly, \cite{wko_59} which has been reported for a variety of examples. \cite{jkr_06}  It describes the softening of phonons with wave vectors that coincide with $\textbf{k}_1 - \textbf{k}_2 \pm \textbf{g}$ where $\mathbf{k}_1$, $\mathbf{k}_2$ are wave vectors of electrons at the Fermi level and $\mathbf{g}$ denotes a reciprocal lattice vector.  The Fermi surface of pristine graphene consists of two equivalent points at $\bar{\text{K}}$ and $\bar{\text{K}}'$, which reflect the tips of the Dirac cones.  Thus, Kohn anomalies are expected at $\bar{\Gamma}$ and $\bar{\text{K}}$. \cite{mla_06}  Indeed, inelastic X-ray data obtained from the highest optical phonon branches of graphite \cite{jma_04} have been interpreted in terms of the Kohn anomaly. \cite{spi_04}  Indications of Kohn anomalies of graphene on Pt(111) have recently been provided. \cite{apo12b}

Here, we show that investigations into the dynamics of graphene on a metal surface reveal subtle aspects of electron-phonon coupling, electron correlations, and the graphene--metal interaction.  To this end the dispersion of all acoustic and optical phonons of graphene on Ir(111) along high symmetry directions of the surface Brillouin zone is presented. The experimental data are in very good agreement with accompanying density functional calculations.  As the main finding we report the weakened Kohn anomaly of the highest optical phonon branch around $\bar{\text{K}}$.  This observation is rationalized in terms of a reduced electron-phonon interaction due to screening of electron correlations in graphene by the metal electron gas.

\begin{figure*}
\includegraphics[bbllx=70,bblly=10,bburx=725,bbury=595,width=110mm]
{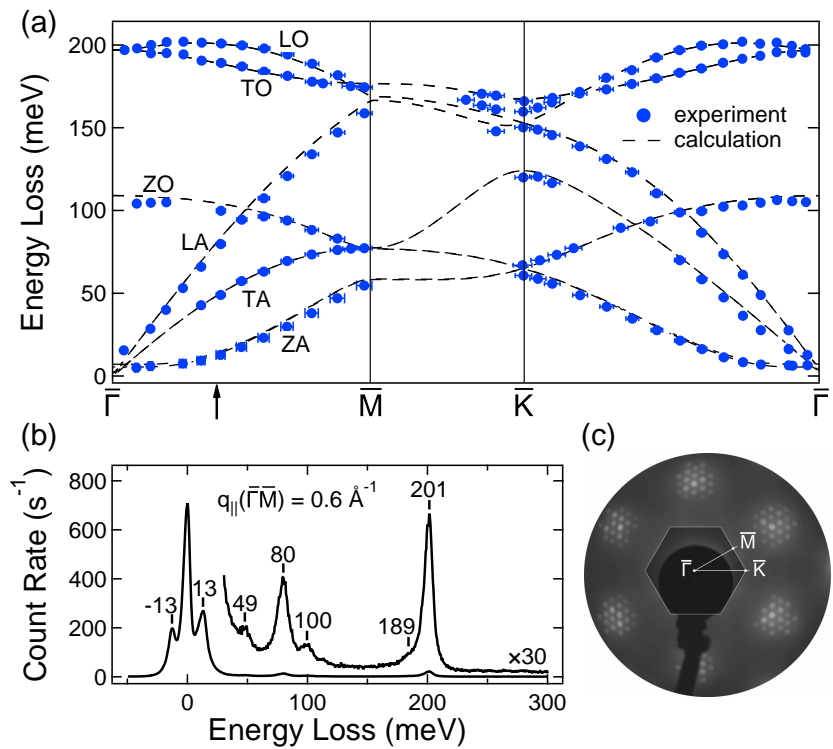}
\caption[fig1]{(color online) (a) Phonon dispersion of graphene on Ir(111).  Dispersion branches of the out-of-plane acoustic (ZA), out-of-plane optical (ZO), transverse acoustic (TA), longitudinal acoustic (LA), transverse optical (TO), and longitudinal optical (LO) phonons are indicated.  Experimental (calculated) data appear as dots (dashed lines).  The distances between the high symmetry points of the surface Brillouin zone of graphene ($\bar{\Gamma}$, $\bar{\text{M}}$, $\bar{\text{K}}$) are $\bar{\Gamma}\bar{\text{M}}=2\pi/(\sqrt{3}a)=1.48\,\text{\AA}^{-1}$ and $\bar{\Gamma}\bar{\text{K}}=4\pi/(3a)=1.71\,\text{\AA}^{-1}$ with $a=2.452\,\text{\AA}$ the lattice constant of graphene on Ir(111).  The arrow indicates the parallel wave vector at which the spectrum in (b) was acquired.  (b) Off-specular electron energy loss spectrum acquired with an impact electron energy of $111\,\text{eV}$ and a wave vector transfer of $0.6\,\text{\AA}^{-1}$ along the $\bar{\Gamma}\bar{\text{M}}$ direction.  Loss features at $13$, $49$, $80$, $100$, $189$, $201\,\text{meV}$ are due to electron scattering from ZA, TA, LA, ZO, TO, LO phonons, respectively.  The peak at $-13\,\text{meV}$ reflects the energy gain of electrons scattered from ZA phonons.  (c) Low-energy electron diffraction pattern of graphene-covered Ir(111) acquired with a kinetic energy of incident electrons of $144\,\text{eV}$.  The surface Brillouin zone with high symmetry points is indicated.  The diffraction spots are due to Ir(111) and the long-range moiré pattern of graphene.}
\label{fig1}
\end{figure*} 

\section{Experiment}

The experiments were performed at room temperature and in ultrahigh vacuum with a base pressure of $10^{-9}\,\text{Pa}$.  Dispersion curves were measured by angle-resolved inelastic electron spectroscopy using an Ibach spectrometer. \cite{hib_93}  The energy resolution was set to $4\,\text{meV}$.  Parallel wave vectors were determined with an accuracy below $0.05\,\text{\AA}^{-1}$.  Ir(111) surfaces were cleaned by Ar$^+$ bombardment and annealing.  Cleanliness of Ir(111) was checked by featureless specular vibrational loss spectra and crystalline order was verified by a sharp low-energy electron diffraction pattern.  A single layer of graphene was prepared via thermal decomposition of C$_2$H$_4$.  Exposure of clean Ir(111) to C$_2$H$_4$ (purity $99.9\,\%$, $5\times 10^{-4}\,\text{Pa}$, $120\,\text{s}$) and subsequent annealing at $1400$--$1500\,\text{K}$ leads to extended highly ordered and singly oriented graphene \cite{hha_11} as revealed by low-energy electron diffraction.

\section{Theory}

The phonon dispersion relations of graphene on Ir(111) were calculated using density functional perturbation theory \cite{abinit,gonze,baroni} with the local density approximation (LDA) to the exchange-correlation functional. \footnote{Local functionals neglect van der Waals forces.  However, the LDA to the exchange functional gives reasonable results for the geometry and for the energies of interlayer phonon modes. \cite{aal_10}}  A periodic unit cell containing three layers of Ir with the graphene honeycomb center located above face-centered cubic Ir(111) sites \footnote{Calculations with the graphene honeycomb center residing at hexagonal close-packed and on-top Ir(111) sites lead to essentially identical results.} on both sides of the slab was used. \cite{aal_10}  The vacuum distance between graphene layers of neighboring supercells was $6\,\text{\AA}$. \footnote{Graphene remains flat in the sandwich configuration.  This configuration is different from a previous report, where the presence of Ir nano-islands on one side of the graphene layer leads to a strong corrugation of graphene. \cite{pfe_08}}  The lattice constant of Ir(111) was adapted to the experimentally determined lattice constant of graphene on Ir(111) ($2.452\,\text{\AA}$ \cite{adi_08}) in order to avoid a large moiré supercell, which would render the {\it ab initio} calculations of phonons unfeasible.  Troullier-Martins pseudopotentials were used with an energy cut-off of $30\,\text{Ha}$.  The first Brillouin zone is sampled by a $12\times12\times1$ $\mathbf{k}$ point grid.  Geometry optimization under these conditions leads to a graphene--Ir distance of $3.64\,\text{\AA}$, which is in reasonable agreement with the experimental value of $3.38\,\text{\AA}$. \cite{cbu_11}  The resulting graphene buckling is less than $0.002\,\text{\AA}$.  In the dynamical matrix the Ir atoms were assigned a large mass to obtain Ir phonon energies close to zero, which facilitates their discrimination from graphene phonons. 

\section{Results and discussion}

Figure \ref{fig1}(a) shows the experimental (dots) and calculated (dashed lines) phonon dispersion of graphene on Ir(111). \footnote{For clarity we do not show graphene phonon branches that are due to backfolding induced by the moiré pattern, nor do we include the dispersion curves of Ir(111).}  Dispersion branches of the out-of-plane acoustic (ZA), out-of-plane optical (ZO), transverse acoustic (TA), longitudinal acoustic (LA), transverse optical (TO), and longitudinal optical (LO) phonons are indicated.  Each data point of the measured phonon dispersion has been extracted from individual off-specular vibration spectra, an example of which is shown in Fig.\,\ref{fig1}(b) for a wave vector transfer of $q_{\parallel}=0.6\,\text{\AA}^{-1}$ along the $\bar{\Gamma}\bar{\text{M}}$ direction.  The low-energy electron diffraction pattern presented in Fig.\,\ref{fig1}(c) demonstrates the quality of the prepared graphene layer.  Each Ir(111) diffraction spot is surrounded by an extended hexagonal array of satellite spots, which is due to the moiré pattern of singly oriented incommensurate graphene. \cite{hha_11}  The absence of additional diffraction spots at angles of $30^\circ$ with respect to the indicated $\bar{\Gamma}\bar{\text{M}}$ directions shows that densely packed graphene orientations are aligned with crystallographic Ir(111) directions. \cite{hha_11}  Experimental and calculated phonon dispersion curves are in excellent agreement.  Moreover, the dispersion curves are very similar to the dispersion curves of pristine graphene \cite{wir_04,lazzeri2008} and graphite, \cite{jma_04,wir_04,lazzeri2008,gru_09} which confirms the reported weak graphene--Ir interaction. \cite{ipl_09}  However, some deviations occur.  Scrutinizing these deviations enables new insights into the graphene--substrate interaction.

\begin{figure}
\includegraphics[bbllx=15,bblly=20,bburx=565,bbury=445,width=85mm]{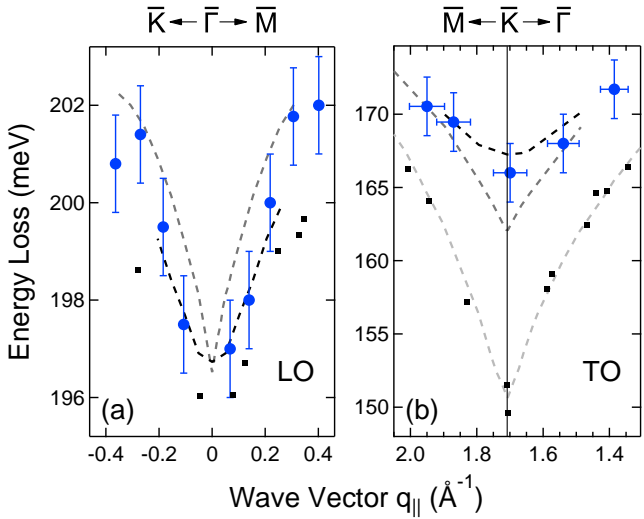}
\caption[fig2]{(color online) Dispersion of (a) the LO phonon in the vicinity of $\bar{\Gamma}$ and of (b) the TO phonon close to $\bar{\text{K}}$. Dots represent our data for graphene on Ir(111), squares are inelastic X-ray data obtained from graphite around $\bar{\Gamma}$ \cite{jma_04} and around $\bar{\text{K}}$, \cite{gru_09} black (gray) dashed lines show LDA density functional calculations for graphene on Ir(111) (pristine graphene), and the light gray dashed line in (b) represents GW calculations for graphite. \cite{gru_09}  The vertical line in (b) indicates the boundary of the surface Brillouin zone at $\bar{\text{K}}$.}
\label{fig2}
\end{figure} 

The first deviation to be discussed concerns the Kohn anomalies of the highest optical phonon branches.  Figure \ref{fig2} shows a close-up view of the dispersion relations of the LO phonon around $\bar{\Gamma}$ [Fig.\,\ref{fig2}(a)] and of the TO phonon around $\bar{\text{K}}$ [Fig.\,\ref{fig2}(b)], both depicted as dots.  Calculations for graphene on Ir(111) appear as black dashed lines.  For comparison, inelastic X-ray scattering data of graphite around $\bar{\Gamma}$ \cite{jma_04} and around $\bar{\text{K}}$ \cite{gru_09} are shown as squares.  Calculated dispersion curves \cite{lazzeri2008} for pristine graphene and graphite have been added as gray and light gray dashed lines, respectively.  Around $\bar{\Gamma}$ [Fig.\,\ref{fig2}(a)] all data sets exhibit good agreement.  In particular, the experimental dispersion of the LO phonon can be very well reproduced by LDA density functional calculations, which neglect the long-range character of the electron-electron interaction.  Consequently, for the LO phonon of graphene on Ir(111) correlation effects play a minor role, which is in line with observations from the LO phonon of pristine graphene and graphite. \cite{lazzeri2008}  The calculations reveal that the LO dispersion curve around $\bar{\Gamma}$ (black dashed line) exhibits a parabolic minimum rather than a kink, which is expected for pristine graphene (gray dashed line).  The shift of the Fermi level due to p-doped graphene \cite{ipl_09} and the temperature-induced broadening of the Fermi-Dirac distribution function are the reasons for this observation.  Apart from these minor deviations, we conclude that the Kohn anomaly of the highest optical phonon branch at $\bar{\Gamma}$ persists in graphene on Ir(111). 

The situation is markedly different for the TO phonon dispersion close to $\bar{\text{K}}$ [Fig.\,\ref{fig2}(b)].  Compared with inelastic X-ray data (squares) and GW calculations for graphite (light gray dashed line) the indentation of the TO dispersion of graphene on Ir(111) is less pronounced.  For instance, the TO phonon energy at $\bar{\text{K}}$ is $\approx 16\,\text{meV}$ ($129\,\text{cm}^{-1}$) higher than observed for graphite.  A hint to the driving mechanism is given by the good agreement between the experimental dispersion data and LDA calculations (black dashed line).  In contrast, for graphite density functional theory within the LDA failed in describing the dispersion of the highest optical phonon branch at $\bar{\text{K}}$. \cite{lazzeri2008,gru_09}  Rather, GW calculations that take electron correlations explicitly into account were required to adequately model experimental data.  Therefore, it appears that correlation effects for the TO phonon of graphene on Ir(111) are less important than observed for pristine graphene and graphite.  Electron correlations in graphene may be reduced by the screening of the Ir(111) electronic system.  To corroborate this scenario the coupling between TO phonons and $\pi$ electrons is analyzed for pristine graphene and for graphene on Ir(111).  The electron-phonon coupling may be obtained from $\langle D^2\rangle=\Delta E^2/(8d^2)$ where $\Delta E$ describes the gap of $\pi$ bands that arises due to the displacement $d$ of C atoms according to the TO phonon pattern at $\bar{\text{K}}$. \cite{lazzeri2008}  Results for $\Delta E$ and $\langle D^2\rangle$ obtained from density functional and GW calculations are summarized in Table \ref{epc}. \footnote{In order to render the calculations feasible, the equivalence of the two C atoms in the graphene unit cell is ensured by attaching three layers of Ir in the face-centered cubic stacking to each side of the graphene sheet.  The GW calculations were performed with YAMBO \cite{yambo} using a $18\times18\times1$ $\mathbf{k}$ grid.} 

\begin{table}
\caption{Band gap ($\Delta E$) and electron-phonon coupling ($\langle D^2\rangle$) for the graphene TO phonon at $\bar{\text{K}}$.  Calculations were performed for pristine graphene and graphene on Ir(111).  The displacement of C atoms is $d=0.53\,\text{pm}$.  Results from LDA and GW calculations are compared.}
\label{epc}
\begin{ruledtabular}
\begin{tabular}{lccc}
 & & pristine & graphene \\
 & & graphene & on Ir(111) \\
\hline
$\Delta E$ (eV)                       & LDA & $0.142$  & $0.142$  \\
                                      & GW  & $0.2158$ & $0.1747$ \\
$\langle D^2\rangle$ (eV$^2$/\AA$^2$) & LDA & $90.11$  & $90.11$ \\
                                      & GW  & $207.88$ & $131.75$
\end{tabular}
\end{ruledtabular}
\end{table}

While density functional calculations give the same results for pristine graphene and graphene on Ir(111), the GW calculations yield an electron-phonon coupling for graphene on Ir(111) that is $\approx 37\,\%$ lower than the value obtained for pristine graphene.  This effect is preponderantly due to the screening of electron correlations in graphene by the metal substrate.  The calculations even underestimate the screening effect since metal intraband contributions to the dielectric function are missing. \cite{caz_10}  To a lesser extent charge transfer between graphene and Ir(111) may be responsible for the reduction of the electron-phonon coupling. \cite{attaccalite2010}  As a result, the metallic substrate reduces the electron-phonon coupling by screening of electron correlations.  In the case of isolated graphene and graphite, correlation effects are responsible for a strong Kohn anomaly at $\bar{\text{K}}$.  Thus, screening of correlation effects by the metallic substrate leads to a reduction of the Kohn anomaly compared to the cases of isolated graphene and graphite.  These observations are in stark contrast to findings reported from graphene on Ni(111), \cite{tai_90,ash_99} where the strong hybridization between graphene $\pi$ bands and Ni $d$ bands causes a destruction of the linear crossing of the $\pi$ and $\pi^*$ bands at the Fermi level, which in turn leads to the elimination of both graphene Kohn anomalies. \cite{aal_10}  For graphene on Ir(111), however, the  Dirac cones remain essentially intact \cite{ipl_09} and both Kohn anomalies persist.  The weakening of the $\bar{\text{K}}$ point Kohn anomaly, however, has not been expected and requires the aforementioned screening mechanism.

A direct measure of the graphene--Ir interaction is the finite energy of the ZA phonon mode at $\bar{\Gamma}$ [$\approx 6\,\text{meV}$, Fig.\,\ref{fig1}(a)], which represents the second deviation from pristine graphene and graphite dispersion curves.  Indeed, at zero wave vector the ZA phonon energy vanishes for pristine graphene \cite{wir_04,lazzeri2008} and graphite. \cite{mmo_07}  In a simple harmonic oscillator model for the C vibrations \cite{aal_10} the observed ZA phonon energy can be translated into a spring constant, $2m\omega^2\approx 3.3\,\text{N}\text{m}^{-1}$ ($m$: C mass, $\omega$: angular frequency of the ZA phonon at $\bar{\Gamma}$).  This spring constant is about a factor $25$ lower than the one obtained for graphene on Ni(111) \cite{aal_10} and about a factor $2$ lower than the spring constant for the interlayer coupling in graphite. \cite{wir_04}  

Our experiments further show that at $\bar{\text{K}}$ the degeneracy of the ZA and ZO (LA and LO) phonons is lifted by $\approx 9\,\text{meV}$ ($\approx 8\,\text{meV}$).  It is tempting to associate this energy splitting to the finite interaction with the substrate.  However, given the weak interaction the calculated ZA--ZO (LA--LO) energy splitting is less than $0.5\,\text{meV}$ ($0.1\,\text{meV}$) and would be even smaller if the lattice mismatch between graphene and Ir(111) was taken into account.  The reason for this discrepancy is unclear at present.  In part, a deviation from the exact $\bar{\Gamma}\bar{\text{K}}$ direction in the experiment may contribute to the observed energy splitting.  Using the low-energy electron diffraction pattern the sample orientation has been adjusted with an accuracy of $\approx 1^\circ$.  Within this accuracy margin phonon modes with a calculated energy difference of $2\,\text{meV}$ are detected around $\bar{\text{K}}$, which, however, is still lower than the observed splitting.

\section{Conclusion}

Graphene on Ir(111) is archetypical in revealing subtle aspects of electron-phonon coupling and electron correlations in graphene.  The weak graphene--metal interaction leaves its clear fingerprints in the phonon dispersion of graphene.  In particular, the modified Kohn anomaly of the highest optical phonon branch is a signature of the reduced electron-phonon coupling, which is induced by the screening of the electron-electron interaction in graphene by the metal electron gas.  This screening efficiently damps correlations effects that are not captured by density functional calculations.  It renders standard local exchange-correlation functionals precise enough for modeling phonon dispersions of graphene on metal surfaces with a weak graphene--metal interaction.  

\acknowledgments
M.~E.\ and J.~K.\ acknowledge funding by the Carl-Zeiss foundation.  A.~M.\ and L.~W.\ acknowledge funding by the French National Research Agency through ANR-09-BLAN-0421-01.  Calculations were performed at the IDRIS supercomputing center, Orsay (Proj.~No.~091827), and at the Tirant Supercomputer of the University of Valencia (group vlc44).

\bibliographystyle{apsrev4-1}

\end{document}